\def\1{\'{\i}}
\def\2{\~n}

\def\b{\beta}

\def\ba{\begin{eqnarray}}
\def\ea{\end{eqnarray}}
\def\be{\begin{equation}}
\def\ee{\end{equation}}
\def\beq{\begin{equation}}
\def\eeq{\end{equation}}

\def\sech{\,{\rm sech}\,}
\def\csch{\,{\rm csch}\,}
\def\sec{\,{\rm sec}\,}
\def\csc{\,{\rm csc}\,}

\documentclass[12pt]{elsarticle}
\usepackage{graphicx}
\usepackage[latin1]{inputenc}
\usepackage{epsfig}
\usepackage{rotating}
\usepackage{amsmath}
\usepackage{slashbox}
\usepackage{setspace}
\usepackage{amsfonts}
\usepackage{dcolumn}
\usepackage{multirow}
\usepackage{url}
\usepackage{color,hyperref}
\hypersetup{colorlinks,breaklinks,
            linkcolor=darkblue,urlcolor=darkblue,
            anchorcolor=darkblue,citecolor=darkblue}
\usepackage{dsfont}

\usepackage{graphicx}

\usepackage{amssymb}
\usepackage{times}

\journal{Elsevier}

\begin{document}

\begin{frontmatter}

\title{Infinite families of shape invariant potentials with $n$ parameters subject to translation}

\author[rvt]{Arturo Ramos\corref{cor1}}
\ead{aramos@unizar.es}

\cortext[cor1]{Corresponding author}

\address[rvt]{
Departamento de An\'alisis Econ\'omico,
Universidad de Zaragoza, \\ Gran V\'{\i}a 2, E-50005 Zaragoza, Spain}

\begin{abstract}
We find new families of shape invariant potentials depending on $n\geq1$ parameters subject to translation by the inclusion of non-trivial invariants. New dependencies of the spectra are found, and it opens the door to the engineering of physical quantities in a novel way. A number of examples are explicitly constructed.
\end{abstract}

\end{frontmatter}

\section{Introduction}

The concept of Shape Invariance in Quantum Mechanics has a long tradition beginning in some of the works of Schr\"odinger himself \cite{Sch40,Sch41,Sch41b} and developed afterwards in a classic work by Infeld and Hull \cite{InfHul51}. These works dealt with exactly solvable problems in Quantum Mechanics. Some years later, Gendenshte\"{\i}n and Krive renamed these cases as shape invariant potentials \cite{Gen83,GenKri85}, although it has been shown later a complete equivalence of the two approaches \cite{CarRam00c}. The list of known shape invariant potentials, summarized for example in the by now classic work \cite{CooKhaSuk95} remained unchanged until key developments by {G\'omez--Ullate}, Kamran and Milson \cite{GomKamMil09} fostered the finding of iso-spectral rational extensions of some of the known shape invariant potentials, first by Quesne \cite{Que08,Que09} and afterwards followed by the same or other researchers, also with related questions \cite{BagQueRoy09,Que11,Que12,
Que12b,OdaSas09,OdaSas10,OdaSas10b,OdaSas10c,OdaSas11,
OdaSas11b,GomGraMil14,Gra11,Gra12,Gra12b,Gra14,
GraBer10,GraBer13,YadKumKhaMan15,YadKhaBagKumMan16}.
In the meanwhile, they have appeared some works showing a compatibility condition that the (rational) extensions should satisfy \cite{Ram11,Ram12} and its equivalence with a group theory condition that appeared in \cite{YadKumKhaMan15} is shown in the later work \cite{RamBagKhaKumManYad17}.
In this last paper the question was open so as to explore the existence of shape invariant potentials (and their extensions) depending on more than one parameter subject to translation, inspired by the article \cite{CarRam00a}.

That research is the subject of the present paper. In fact, the classic work of Infeld and Hull \cite{InfHul51} dealt with a series in the parameter subject to translation which was unable to detect the existence of the rational extensions known much later as we have indicated. Likewise, \cite{CarRam00c} dealt with series expansions in the parameters subject to translation, and essentially only one solution was found, being unable to find more solutions. Both approaches revealed to be not the most appropriate, as the rational extensions appeared much later by other means. And in this work we find an infinite number of shape invariant potentials depending on $n\geq1$ parameters subject to translation.

The rest of the paper is organized as follows. Section~\ref{equiv} summarizes the main idea in \cite{CarRam00a}. In Section~\ref{nprop} we propose the new solutions. In Section~\ref{specsol} we
find the basic specific solutions obeying the general setting. In Section~\ref{preknown} we discuss explicitly the relation of the approach with the previously known shape invariant cases in two parameters. In Section~\ref{ratext} we briefly describe the use of the $n\geq1$ parameters subject to translation and rational extensions. In Section~\ref{moregen} we discuss the generalization to $n\geq1$ parameters of the cases of shape invariant potentials whose rational extensions are not known to date. Finally, in the last Section we offer  some conclusions.

\section{The original idea}\label{equiv}

In \cite{CarRam00a} it is proposed the following form of the superpotential depending on $m_1,m_2,\dots,m_n$ parameters subject to simultaneous translation $m_i\rightarrow m_i-1$, $i=1,\dots,n$:
$$
k(x;m_1,\dots,m_n)=g_0(x)+\sum_{i=1}^n m_i g_i(x)
$$
where $g_0(x),g_1(x),\dots,g_n(x)$ are functions to be determined to satisfy the shape invariance condition (in an obvious notation)
\begin{equation}
k^2(x;m_i+1)-k^2(x;m_i)+k'(x;m_i+1)+k'(x;m_i)=L(m_i)-L(m_i+1)
\label{sicond}
\end{equation}
Upon substitution and selecting the coefficients of the different expressions in $m_i$, we have that the following system of equations have to be satisfied:
\begin{eqnarray}
  g_j^{\prime}+g_j\sum_{i=1}^n g_i &=& c_j,\,\quad\, j=1,\dots,n \label{system}\\
   g_0^{\prime}+g_0\sum_{i=1}^n g_i &=& c_0\label{system2}
\end{eqnarray}
where $c_0,c_1,\dots,c_n$ are constants.
This system is difficult to be solved directly, as \cite{CooKhaSuk95} pointed out. A way to solve it was found in \cite{CarRam00a} and is to take barycentre coordinates of the $g_i$:
\begin{eqnarray}
  g_{{\rm cm}}(x)&=&\frac{1}{n}\sum_{i=1}^n g_i(x) \nonumber\\
  v_i(x)&=&g_i(x)-g_{{\rm cm}}(x)\,,\quad i=1,\dots,n \nonumber\\
  c_{{\rm cm}}&=&\frac{1}{n}\sum_{i=1}^n c_i\nonumber
\end{eqnarray}
Then, the system (\ref{system}), (\ref{system2}) decouples as follows (note that $v_1(x)=-\sum_{j=2}^n v_j(x)$):
\begin{eqnarray}
  (n g_{{\rm cm}})'+(n g_{{\rm cm}})^2 &=& n c_{{\rm cm}} \label{eqcm}\\
  v_j^{\prime}+v_j n g_{{\rm cm}} &=& c_j-c_{{\rm cm}},\quad j=2,\dots,n \label{relcor}\\
g_0^{\prime}+g_0 n g_{{\rm cm}} &=& c_0 \label{eqg0}
\end{eqnarray}
out of which we can distinguish a Riccati equation with constant coefficients for $ng_{{\rm cm}}$ and once solved, linear equations for the rest of functions, and this becomes a standard problem, very well known in the literature \citep{InfHul51,CarRam00c}.

What interests more us is that in terms of the
functions $n g_{{\rm cm}}$, $v_j$ the initial superpotential can be written in the following way:
\begin{equation}
k(x;m_i)=g_0(x)+\sum_{j=2}^n(m_j-m_1)v_j
+\left(\frac{1}{n}\sum_{i=1}^n m_i\right)n g_{{\rm cm}}(x)
\end{equation}
We can observe that under the change $m_i\rightarrow m_i-1$, $i=1,\dots,n$, the quantity
$$
M=\frac{1}{n}\sum_{i=1}^n m_i
$$
changes as $M\rightarrow M-1$ and that $(m_j-m_1)$ are \emph{invariant} for all $j=2,\dots,n$.

\section{A new proposal}\label{nprop}

Inspired by the previous idea, we propose a new form for the superpotential:
\begin{equation}
k(x;m_i)=\sum_{j=1}^r I_j v_j(x)+M\,G(x)
\label{kMG}
\end{equation}
where $M$ has been defined in the preceding Section, $r$ is a positive integer, and $I_j$ are
expressions in the $m_i$, $i=1,\dots,n$ \emph{invariant} under translation, that is
$$
I_j(m_1-k,m_2-k,\dots,m_n-k)=I_j(m_1,m_2,\dots,m_n)
$$
for all $j=1,\dots,r$ and for all $k\in\mathds{N}$. There is an infinite number of such invariants, and they can be  linear or non-linear in $m_i$.
This setting allows in particular \emph{invariants in only one parameter}, that is, if $n=1$, we can have
$$
I_j(m_1-k)=I_j(m_1),\quad j=1,\dots,r
$$
for all $k\in\mathds{N}$.
That is, periodic functions in one parameter $m_1$ with period $1$ will do the job.
The $G(x)$ and $v_j(x)$ appearing in (\ref{kMG}) are required to satisfy the following system of differential equations:
\begin{eqnarray}
  G'+G^2 &=&\alpha \label{eqRic}\\
  v_j^{\prime}+v_j G &=&\beta_j,\quad j=1,\dots,r \label{eqLin}
\end{eqnarray}

Then, let us check that the shape invariance condition is met (in order to avoid excessive notation we will assume in what follows summation in the repeated index in an obvious way):
\begin{eqnarray}
&& k^2(x;m_i+1)-k^2(x;m_i)+k'(x;m_i+1)+k'(x;m_i)=\nonumber\\
&&(I_j v_j+(M+1)G)^2-(I_j v_j+MG)^2
+(I_jv_j+(M+1)G)'+(I_jv_j+MG)'\nonumber\\
&&=(2M+1)(G^2+G')+2 I_j(v_j G+v_j')
=(2M+1)\alpha+2\beta_jI_j=L(m_i)-L(m_i+1)=R(m_i)
\end{eqnarray}

Or, constructing the following partner potentials and remainder $R$:
\begin{eqnarray}
V(x;m_i)&=&k(x;m_i)^2-k'(x;m_i)\nonumber\\
&=&M(M+1)G^2+(2M+1)I_jv_j G+(I_jv_j)^2-M\alpha-\beta_jI_j\nonumber\\
\tilde{V}(x;m_i)&=&k(x;m_i)^2+k'(x;m_i)\nonumber\\
&=&M(M-1)G^2+(2M-1)I_jv_jG
+(I_jv_j)^2+M\alpha+\beta_jI_j\nonumber\\
R(m_i)&=&(2M+1)\alpha+2\beta_jI_j\nonumber
\end{eqnarray}
Then, it is immediate to check that it holds
$$
\tilde{V}(x;m_i)=V(x;m_i-1)+R(m_i-1)
$$
so it is satisfied the shape invariance condition.

\section{Specific solutions}\label{specsol}

The solutions of (\ref{eqRic}) are very well-known, that are summarized for example in \cite{InfHul51,EngQue91,CarRam00c}.

In fact, if $\alpha>0$, by a convenient re-scaling of the variable can be carried into $\alpha=1$, and the solutions can be reduced to any of the following four types:
\begin{eqnarray}
  G(x)&=&\tanh(x) \label{soltanh}\\
  G(x)&=&\coth(x) \label{solcoth}\\
  G(x)&=&1 \label{solc}\\
  G(x)&=&-1\label{solmc}
\end{eqnarray}
The corresponding solutions of (\ref{eqLin}) can be written as, respectively:
\begin{eqnarray}
   v_j(x)&=&\beta_j\tanh(x)+d_j\sech(x) \\
  v_j(x)&=&\beta_j\coth(x)-d_j\csch(x) \\
  v_j(x)&=&\beta_j-d_j{\rm e}^{-x} \\
  v_j(x)&=&-\beta_j-d_j{\rm e}^{x}
\end{eqnarray}
where $\beta_j,d_j$, $j=1,\dots,r$ are real constants.

If $\alpha=0$ the following two basic types can be obtained:
\begin{eqnarray}
  G(x)&=&1/x \label{Gcoul}\\
  G(x) &=&0 \label{Goscil}
\end{eqnarray}
and the corresponding solutions to (\ref{eqLin}) can be written, respectively, as:
\begin{eqnarray}
  v_j(x) &=&\frac{\beta_j}{2}x+\frac{d_j}{x}\label{vjCoul}\\
  v_j(x) &=& \beta_j x+d_j\label{vjOscil}
\end{eqnarray}
where $\beta_j,d_j$, $j=1,\dots,r$ are real constants.

Likewise, if $\alpha<0$, by a convenient re-scaling of the variable, can be carried into the case $\alpha=-1$; they can be found two types of basic \emph{real} solutions:
\begin{eqnarray}
  G(x)&=&-\tan(x) \label{soltan}\\
  G(x)&=&\cot(x)\label{solcot}
\end{eqnarray}
and the corresponding solutions to (\ref{eqLin}) can be written, respectively, as:
\begin{eqnarray}
  v_j(x) &=& \beta_j\tan(x)-d_j\sec(x) \\
  v_j(x) &=& -\beta_j\cot(x)+d_j\csc(x)
\end{eqnarray}
where $\beta_j,d_j$, $j=1,\dots,r$ are real constants.

The superpotentials so found, according to (\ref{kMG}) can, respectively, be written as:
\begin{eqnarray}
k(x;m_i)&=&\left(M+I_j\beta_j\right)\tanh(x)+I_j d_j\sech(x)\label{case1}\\
k(x;m_i)&=&\left(M+I_j\beta_j\right)\coth(x)-I_j d_j\csch(x)\label{case2}\\
k(x;m_i)&=&\left(M+I_j\beta_j\right)-I_j d_j{\rm e}^{-x}\label{case3}\\
k(x;m_i)&=&-\left(M+I_j\beta_j\right)-I_j d_j{\rm e}^{x}\label{case4}\\
k(x;m_i)&=&\frac{1}{2}\beta_jI_jx+\left(M+d_j I_j\right)\frac{1}{x}\label{case5}\\
k(x;m_i)&=&I_j\beta_j x+d_j I_j\label{case6}\\
k(x;m_i)&=&-\left(M-I_j\beta_j\right)\tan(x)-I_j d_j\sec(x)\label{case7}\\
k(x;m_i)&=&\left(M-I_j\beta_j\right)\cot(x)+I_j d_j\csc(x)\label{case8}
\end{eqnarray}

We will write these superpotentials in terms of three new quantities $\epsilon,\rho,\beta$.
In the cases (\ref{case1}), (\ref{case2}), (\ref{case3}), (\ref{case4}) we will set $\epsilon=M+\beta_j I_j$, $\rho=d_j I_j$. In the case (\ref{case5}) we will set $\epsilon=M+d_jI_j$ and $\rho=\frac{1}{2}\beta_jI_j$. In the case (\ref{case6}) we will set $\beta=\beta_j I_j$ and $\rho=d_j I_j$. And in the cases (\ref{case7}) and (\ref{case8}) we will set $\epsilon=M-\beta_j I_j$ and $\rho=d_j I_j$.

Let us describe in what follows the corresponding superpotential, partner potentials, remainder of the shape invariance condition, and eigenenergies and normalized eigenstates of the potential $V(x;\epsilon,\rho)$ or $V(x;\beta,\rho)$ for each case.

\subsection{Case (\ref{case1}) (Scarf 2 type)}

We have set $\epsilon=M+\beta_j I_j$, $\rho=d_j I_j$. We have, for $\epsilon>0$:
\begin{eqnarray}
k(x;\epsilon,\rho)&=&\epsilon\tanh(x)+\rho\sech(x),\quad x\in(-\infty,\infty)\nonumber\\
V(x;\epsilon,\rho)&=&\epsilon^2 \tanh^2(x)+\rho(2\epsilon+1)\tanh(x)\sech(x)+(\rho^2-\epsilon)\sech^2(x)\nonumber\\
\tilde{V}(x;\epsilon,\rho)&=&\epsilon^2 \tanh^2(x)+\rho(2\epsilon-1)\tanh(x)\sech(x)+(\rho^2+\epsilon)\sech^2(x)\nonumber\\
R(\epsilon,\rho)&=&2\epsilon+1\\
E_k&=&(2\epsilon-k)k\nonumber\\
\zeta_k(x;\epsilon,\rho)&=&
2^{\epsilon-1/2}\frac{|\Gamma\left(1/2+\epsilon-k-i\rho\right)|}{\sqrt{\pi}\sqrt{\Gamma(2(\epsilon-k))}}k!i^k a_k(\epsilon)
{\rm e}^{-\rho \arctan(\sinh(x))}(\cosh(x))^{-\epsilon}
  P_k^{(-1/2-\epsilon+i \rho,-1/2-\epsilon-i \rho)}(-i\sinh(x))\nonumber\\
\end{eqnarray}
where
\begin{equation}
a_k(\epsilon)=\left\{\begin{array}{ll}
1, & k=0\\
\frac{a_{k-1}(\epsilon-1)}{\sqrt{(2\epsilon-k)k}}, & k>0
\end{array}\right.\label{eqa}
\end{equation}
and $P_k^{(a,b)}(x)$ is a (ordinary) Jacobi polynomial of degree $k$, $\Gamma(\cdot)$ is the usual Gamma function, and $i$ is the imaginary unit (from the context it should not be considered as a summation index).

\subsection{Case (\ref{case2}) (P\"oschl-Teller type)}

We have set $\epsilon=M+\beta_j I_j$, $\rho=d_j I_j$. We have for $\epsilon-\rho<1/2,\epsilon>0$:
\begin{eqnarray}
k(x;\epsilon,\rho)&=&\epsilon\coth(x)-\rho\csch(x),\quad x\in(0,\infty)\nonumber\\
V(x;\epsilon,\rho)&=&\epsilon^2 \coth^2(x)-\rho(2\epsilon+1)\coth(x)\csch(x)
+(\rho^2+\epsilon)\csch^2(x)\nonumber\\
\tilde{V}(x;\epsilon,\rho)&=&\epsilon^2 \coth^2(x)-\rho(2\epsilon-1)\coth(x)\csch(x)
+(\rho^2-\epsilon)\csch^2(x)\nonumber\\
R(\epsilon,\rho)&=&2\epsilon+1\nonumber\\
E_k&=&-k(k-2\epsilon)\nonumber\\
\zeta_k(x;\epsilon,\rho)&=&2^{\epsilon}
\sqrt{\frac{\Gamma(1/2-k+\epsilon+\rho)}
{\Gamma(2(\epsilon-k))\Gamma(1/2+k-\epsilon+\rho)}}
k!b_k(\epsilon)\nonumber\\
&&(\cosh(x)-1)^{(-\epsilon+\rho)/2}
(\cosh(x)+1)^{-(\epsilon+\rho)/2}
P_k^{(-1/2-\epsilon-\rho,-1/2-\epsilon+\rho)}(-\cosh(x))\nonumber
\end{eqnarray}
where
\begin{equation}
b_k(\epsilon)=\left\{\begin{array}{ll}
1, & k=0\\
\frac{b_{k-1}(\epsilon+1)}{\sqrt{k(2\epsilon-k)}}, & k>0
\end{array}\right.
\end{equation}

\subsection{Case (\ref{case3}) (Morse type)}

We have set $\epsilon=M+\beta_j I_j$, $\rho=d_j I_j$. We have for $\epsilon>0$, $\rho>0$:
\begin{eqnarray}
k(x;\epsilon,\rho)&=&\epsilon-\rho{\rm e}^{-x},\quad x\in(-\infty,\infty)\nonumber\\
V(x;\epsilon,\rho)&=&\rho^2{\rm e}^{-2x}-\rho(2\epsilon+1){\rm e}^{-x}+\epsilon^2\nonumber\\
\tilde{V}(x;\epsilon,\rho)&=&\rho^2{\rm e}^{-2x}-\rho(2\epsilon-1){\rm e}^{-x}+\epsilon^2\nonumber\\
R(\epsilon,\rho)&=&2\epsilon+1\nonumber\\
E_k&=&(2\epsilon-k)k\nonumber\\
\zeta_k(x;\epsilon,\rho)&=&(-1)^k 2^{\epsilon-k} {\rm e}^{kx}\rho^{\epsilon-k}a_k(\epsilon)k!
\frac{\exp(-\rho {\rm e}^{-x}){\rm e}^{-\epsilon x}}{\sqrt{\Gamma(2(\epsilon-k))}}
L_k^{(2\epsilon-2k)}(2\rho {\rm e}^{-x})
\end{eqnarray}
where $a_k(\epsilon)$ is given by (\ref{eqa}) and $L_k^{(a)}(x)$ is a Laguerre polynomial of degree $k$.

\subsection{Case (\ref{case4})}

We have set $\epsilon=M+\beta_j I_j$, $\rho=d_j I_j$. We have for $\epsilon>0$, $\rho<0$:
\begin{eqnarray}
k(x;\epsilon,\rho)&=&-\epsilon-\rho{\rm e}^x,\quad x\in(-\infty,\infty)\nonumber\\
V(x;\epsilon,\rho)&=&\rho^2 {\rm e}^{2x}+\rho(2\epsilon+1){\rm e}^{x}+\epsilon^2\nonumber\\
\tilde{V}(x;\epsilon,\rho)&=&\rho^2 {\rm e}^{2x}+\rho(2\epsilon-1){\rm e}^{x}+\epsilon^2\nonumber\\
R(\epsilon,\rho)&=&2\epsilon+1\nonumber\\
E_k&=&(2\epsilon-k)k\nonumber\\
\zeta_k(x;\epsilon,\rho)&=&(-1)^k 2^{\epsilon-k} {\rm e}^{-kx}\rho^{\epsilon-k}a_k(\epsilon)k!
\frac{\exp(\rho {\rm e}^{x}){\rm e}^{\epsilon x}}{\sqrt{\Gamma(2(\epsilon-k))}}
L_k^{(2\epsilon-2k)}(-2\rho {\rm e}^{x})
\end{eqnarray}
where $a_k(\epsilon)$ is given again by (\ref{eqa}).

\subsection{Case (\ref{case5}) (Radial harmonic oscillator type)}

We have set $\epsilon=M+d_jI_j$ and $\rho=\frac{1}{2}\beta_jI_j$.
We have for $\epsilon<\frac{1}{2}$, $\rho>0$:
\begin{eqnarray}
k(x;\epsilon,\rho)&=&\frac{\epsilon}{x}+\rho x,\quad x\in(0,\infty)\nonumber\\
V(x;\epsilon,\rho)&=&\rho^2 x^2+\rho(2\epsilon-1)
+\frac{\epsilon(\epsilon+1)}{x^2}\nonumber\\
\tilde{V}(x;\epsilon,\rho)&=&\rho^2 x^2+\rho(2\epsilon+1)
+\frac{\epsilon(\epsilon-1)}{x^2}\nonumber\\
R(\epsilon,\rho)&=&4 \rho\nonumber\\
E_k&=&4 \rho k\nonumber\\
\zeta_k(x;\epsilon,\rho)&=&
\sqrt{\frac{2\rho^{1/2+k-\epsilon}}
{\Gamma(1/2+k-\epsilon)}}k!(-2)^k c_k(\epsilon)
\exp(-\rho x^2/2)x^{-\epsilon}
L_k^{(-1/2-\epsilon)}(\rho x^2)
\end{eqnarray}
where
\begin{equation}
c_k(\epsilon)=\left\{\begin{array}{ll}
1, & k=0\\
\frac{c_{k-1}(\epsilon-1)}{\sqrt{4\rho k}}, & k>0
\end{array}\right.
\end{equation}

\subsection{Case (\ref{case6}) (Harmonic oscillator type)}

We have set $\beta=\beta_j I_j$ and $\rho=d_j I_j$. We have, for $\beta>0$:
\begin{eqnarray}
k(x;\beta,\rho)&=&\beta x+\rho,\quad x\in(-\infty,\infty)\nonumber\\
V(x;\beta,\rho)&=&\rho^2+2\rho\beta x+\beta(\beta x^2-1)\nonumber\\
\tilde{V}(x;\beta,\rho)&=&\rho^2+2\rho\beta x+\beta(\beta x^2+1)\nonumber\\
R(\beta,\rho)&=&2\beta\nonumber\\
E_k&=&2\beta k\nonumber\\
\zeta_k(x;\beta,\rho)&=&
\left(\frac{\beta}{\pi}\right)^{1/4}
\left(\frac{1}{\sqrt{k!2^k}}\right)
\exp\left(-\frac{\beta}{2}\left(x+\frac{\rho}{\beta}\right)^2\right)
H_k\left(\sqrt{\beta}\left(x+\frac{\rho}{\beta}\right)\right)
\end{eqnarray}
where $H_k(x)$ is a (ordinary) Hermite polynomial of degree $k$.

\subsection{Case (\ref{case7}) (Scarf 1 type)}

We have set $\epsilon=M-\beta_j I_j$ and $\rho=d_j I_j$.
We have, for
$\epsilon<\frac{1}{2},\frac{1}{2}(2\epsilon-1)
<\rho<\frac{1}{2}(1-2
\epsilon)$,
\begin{eqnarray}
k(x;\epsilon,\rho)&=&-\epsilon\tan(x)-\rho\sec(x),\quad x\in(-\pi/2,\pi/2)\nonumber\\
V(x;\epsilon,\rho)&=&\epsilon^2 \tan^2(x)+\rho(2\epsilon+1)\tan(x)\sec(x)+(\rho^2+\epsilon)\sec^2(x)\nonumber\\
\tilde{V}(x;\epsilon,\rho)&=&\epsilon^2 \tan^2(x)+\rho(2\epsilon-1)\tan(x)\sec(x)+(\rho^2-\epsilon)\sec^2(x)\nonumber\\
R(\epsilon,\rho)&=&-2\epsilon-1\nonumber\\
E_k&=&(k-2\epsilon)k\nonumber\\
\zeta_k(x;\epsilon,\rho)&=&
2^{\epsilon}k!\sqrt{\frac{\Gamma(1+2k-2\epsilon)}{\Gamma(1/2+k-\epsilon-\rho)\Gamma(1/2+k-\epsilon+\rho)}}
d_k(\epsilon)\nonumber\\
&&(1-\sin(x))^{-(\epsilon+\rho)/2}
(1+\sin(x))^{-(\epsilon-\rho)/2}
P_k^{(-1/2-\epsilon-\rho,-1/2-\epsilon+\rho)}(\sin(x))\nonumber\\
\end{eqnarray}
where
\begin{equation}
d_k(\epsilon)=\left\{\begin{array}{ll}
1, & k=0\\
\frac{d_{k-1}(\epsilon-1)}{\sqrt{k(k-2\epsilon)}}, & k>0
\end{array}\right.
\label{eqd}
\end{equation}

\subsection{Case (\ref{case8})}

We have set $\epsilon=M-\beta_j I_j$ and $\rho=d_j I_j$.
We have, for
$\epsilon<\frac{1}{2},\frac{1}{2}(2\epsilon-1)
<\rho<\frac{1}{2}(1-2\epsilon)$,
\begin{eqnarray}
k(x;\epsilon,\rho)&=&\epsilon\cot(x)+\rho\csc(x),\quad x\in(0,\pi)\nonumber\\
V(x;\epsilon,\rho)&=&\epsilon^2 \cot^2(x)+\rho(2\epsilon+1)\cot(x)\csc(x)+(\rho^2+\epsilon)\csc^2(x)\nonumber\\
\tilde{V}(x;\epsilon,\rho)&=&\epsilon^2 \cot^2(x)+\rho(2\epsilon-1)\cot(x)\csc(x)+(\rho^2-\epsilon)\csc^2(x)\nonumber\\
R(\epsilon,\rho)&=&-2\epsilon-1\nonumber\\
E_k&=&(k-2\epsilon)k\nonumber\\
\zeta_k(x;\epsilon,\rho)&=&
2^{\epsilon}k!\sqrt{\frac{\Gamma(1+2k-2\epsilon)}{\Gamma(1/2+k-\epsilon-\rho)\Gamma(1/2+k-\epsilon+\rho)}}
d_k(\epsilon)\nonumber\\
&&(1-\cos(x))^{-(\epsilon+\rho)/2}
(1+\cos(x))^{-(\epsilon-\rho)/2}
P_k^{(-1/2-\epsilon-\rho,-1/2-\epsilon+\rho)}(\cos(x))\nonumber\\
\end{eqnarray}
where $d_k(\epsilon)$ is given again by (\ref{eqd}).

\subsection{Non-trivial examples in one and three parameters}

The first non-trivial example might be to consider the case (\ref{case7}) with only one invariant, so $r=1$, only one parameter, so $n=1$, and then
\begin{eqnarray}
M&=&m_1\nonumber\\
I_1(m_1)&=&\sin^2(2\pi m_1)
+\cos(2\pi m_1)+1\nonumber\\
\epsilon&=&m_1-\beta_1 I_1\nonumber\\
\rho&=&d_1 I_1\nonumber
\end{eqnarray}
where $\beta_1$, $d_1$, $m_1$ are chosen so as to ensure
$\epsilon<\frac{1}{2},\frac{1}{2}(2\epsilon-1)
<\rho<\frac{1}{2}(1-2
\epsilon)$.
Then, the spectrum of the potential is
$$
E_k=(k-2\epsilon)k
=(k-2(m_1-\beta_1(\sin^2(2\pi m_1)
+\cos(2\pi m_1)+1)))k
$$
for some $k$, which depends in a new non-trivial way on $m_1$.

Another specific example of the case (\ref{case7}) could be as follows. Let us choose $r=1$ (only one invariant), $n=3$ (three parameters $m_1,m_2,m_3$),
and
\begin{eqnarray}
M&=&\frac{1}{3}(m_1+m_2+m_3)\nonumber\\
I_1(m_1,m_2,m_3)&=&
\sin(2\pi M)+\sin^2(M-m_1)+\sin^2(M-m_2)+\cos^2(M-m_3)\nonumber\\
\epsilon&=&M-\beta_1 I_1\nonumber\\
\rho&=&d_1 I_1\nonumber
\end{eqnarray}
where $\beta_1$, $d_1$, $m_1,m_2,m_3$ are chosen so as to ensure
$\epsilon<\frac{1}{2},\frac{1}{2}(2\epsilon-1)
<\rho<\frac{1}{2}(1-2
\epsilon)$.
Then, the spectrum of the potential is
$$
E_k=(k-2\epsilon)k
=k\left(k-2M+2
\beta_1\left(\sin(2\pi M)+\sin^2(M-m_1)+\sin^2(M-m_2)+\cos^2(M-m_3)
\right)\right)
$$
for some $k$, which depends in a new non-trivial way on $M,m_1,m_2,m_3$.

\section{Previously known potentials with two parameters}\label{preknown}

There are two cases of previously well-known superpotentials in two parameters subject to translation (see, e.g., \cite{CooKhaSuk95,CooKhaSuk01,GanMalRas11}), namely
\begin{eqnarray}
  k(x;m_1,m_2) &=& m_1\tanh(x)+ m_2\coth(x),\quad {\rm (P\ddot{o}schl-Teller\ II)}\label{ksi21}\\
  k(x;m_1,m_2) &=& -m_1\tan(x)+m_2\cot(x),\quad {\rm (P\ddot{o}schl-Teller\ I)}\label{ksi22}
\end{eqnarray}
Let us show that both can be understood in the previous framework.

In fact, for (\ref{ksi21}) we set $r=1$ and
\begin{eqnarray}
  G(x) &=& 2\coth(2x) \nonumber\\
  v_1(x) &=& 2\csch(2x)\nonumber\\
  M&=&\frac{1}{2}(m_1+m_2)\nonumber\\
  I_1(m_1,m_2)&=&\frac{1}{2}(m_2-m_1)\nonumber
\end{eqnarray}
that satisfy
\begin{eqnarray}
  G'(x)+G(x)^2 &=&4 \nonumber\\
  v_1'(x)+v_1(x)G(x)&=&0\nonumber
\end{eqnarray}
and then, according to (\ref{kMG}),
$$
k(x;m_1,m_2)=M G(x)+I_1(m_1,m_2)v_1(x)
=\frac{1}{2}(m_1+m_2)2\coth(2x)+\frac{1}{2}(m_2-m_1)2\csch(2x)
=m_1\tanh(x)+m_2\coth(x)
$$
This superpotential is of the type (\ref{case2}) re-scaling $x\rightarrow 2x$, $\beta_1=0$ and $d_1=-1$.
Since $\beta_1=0$, the spectrum does not depend on the invariant $I_1(m_1,m_2)$ but only on $M=\frac{1}{2}(m_1+m_2)=\epsilon$. A different potential, with the same spectrum, can be obtained setting, for example, $I_1(m_1,m_2)={\rm e}^{m_2-m_1}+1$ instead of $I_1(m_1,m_2)=\frac{1}{2}(m_2-m_1)$, and with this modification the superpotential no longer takes the form (\ref{ksi21}).

A similar thing can be said about (\ref{ksi22}).
In fact, setting $r=1$ and
\begin{eqnarray}
  G(x) &=& 2\cot(2x) \nonumber\\
  v_1(x) &=& 2\csc(2x)\nonumber\\
  M&=&\frac{1}{2}(m_1+m_2)\nonumber\\
  I_1(m_1,m_2)&=&\frac{1}{2}(m_2-m_1)\nonumber
\end{eqnarray}
that satisfy
\begin{eqnarray}
  G'(x)+G(x)^2 &=&-4 \nonumber\\
  v_1'(x)+v_1(x)G(x)&=&0 \nonumber
\end{eqnarray}
and then, according to (\ref{kMG}),
$$
k(x;m_1,m_2)=M G(x)+I_1(m_1,m_2)v_1(x)
=\frac{1}{2}(m_1+m_2)2\cot(2x)+\frac{1}{2}(m_2-m_1)2\csc(2x)
=-m_1\tan(x)+m_2\cot(x)
$$
This superpotential is of the type (\ref{case8}) re-scaling $x\rightarrow 2x$, and putting $\beta_1=0$, $d_1=1$.
Since $\beta_1=0$, the spectrum does not depend again on the invariant $I_1(m_1,m_2)$ but only on $M=\frac{1}{2}(m_1+m_2)=\epsilon$. A different potential, with the same spectrum, can be obtained setting, for example, $I_1(m_1,m_2)=\ln((m_2-m_1)^2+1)$ instead of $I_1(m_1,m_2)=\frac{1}{2}(m_2-m_1)$, and with this modification the superpotential no longer takes the form (\ref{ksi22}).

\section{Rational extensions}\label{ratext}

In the articles \cite{Ram11,Ram12} it has been established a compatibility condition that the
previously known rational extensions \citep{Que08,Que09,BagQueRoy09,BouGanMal11,OdaSas09,OdaSas10,OdaSas10b,OdaSas10c,OdaSas11,OdaSas11b} should satisfy.
Namely, if $W_0(x;m)=k_0(x)+m k_1(x)$ is a super-potential of the classical type \cite{InfHul51,CarRam00c},
where
\begin{eqnarray}
  k_1'+k_1^2&=&\alpha \label{eqalpha}\\
  k_0'+k_0k_1&=&\beta \label{eqbeta}
\end{eqnarray}
being $\alpha,\beta$ constants,
then the extensions $W(x;m)=W_0(x;m)+W_{1+}(x;m)-W_{1-}(x;m)$,
define shape invariant potentials with $m$ subject to translation $m\rightarrow m-1$ if and only if they are satisfied the following two conditions \cite{Ram11,Ram12}:
\begin{eqnarray}
   &&W_{1+}^2+W_{1+}'+W_{1-}^2+W_{1-}'+2W_0W_{1+}-2W_0W_{1-}-2W_{1+}W_{1-}=h\label{cond1}\\
  && W_{1-}(x;m)=W_{1+}(x;m-1)\label{cond2}
\end{eqnarray}
where (\ref{cond1}) is evaluated on $(x;m)$ or $(x;m-1)$ and $h$ is a function of $x$ only. The appearance of $h$ is due to the existence of symmetries \citep{Ram12} of $W_{1+},W_{1-}$; in fact, it can be added to them the same  arbitrary (differentiable) function of $x$ only.
When more than one parameter subject to translation are involved, it is easy to see that the symmetries extend to the adding of an arbitrary differentiable function $f(x;J_k)$ to $W_{1+}$ and $W_{1-}$, where $J_k$, $k=1,\dots,s$ are also invariants under the change $m_i\rightarrow m_i-1$, $i=1,\dots,n$. The invariants $J_k$ do not need to coincide with the previous $I_j$.

Let us see how a number of cases fit in our approach, extracted and adapted {}from the cited literature.
All of the following cases satisfy (\ref{cond1}) evaluated at $(x;\epsilon,\rho)$,  $(x;\epsilon,\rho,\ell)$ or $(x;\epsilon,\ell)$ where in addition $h(x)=2f'(x;J_k)$ and (\ref{cond2}) with the change in the notation $W_{1-}(x;\epsilon,\rho)=W_{1+}(x;\epsilon-1,\rho)$, $W_{1-}(x;\epsilon,\rho,\ell)=W_{1+}(x;\epsilon-1,\rho,\ell)$
or $W_{1-}(x;\epsilon,\ell)=W_{1+}(x;\epsilon-1,\ell)$, depending on the case.

\begin{enumerate}

\item We have
\begin{eqnarray}
W_0(x;\epsilon,\rho)&=&\epsilon\coth(x)-\rho\csch(x)\nonumber\\
W_{1+}(x;\epsilon,\rho)&=&-\frac{2\rho\sinh(x)}{2\epsilon+1-2\rho\cosh(x)}+f(x;J_k)\nonumber\\
W_{1-}(x;\epsilon,\rho)&=&-\frac{2\rho\sinh(x)}{2\epsilon-1-2\rho\cosh(x)}+f(x;J_k)\nonumber\\
\end{eqnarray}
where $\epsilon=M+\beta_jI_j$, $\rho=d_jI_j$, $M=\frac{1}{n}\sum_{i=1}^n m_i$, $\beta_j,d_j$ are constants and $I_j,J_k$ are invariants in the $m_i$ under the change $m_i\rightarrow m_i-1$, $i=1,\dots,n$.

\item We have
\begin{eqnarray}
W_0(x;\epsilon,\rho,\ell)&=&\epsilon\coth(x)-\rho\csch(x)\nonumber\\
W_{1+}(x;\epsilon,\rho,\ell)&=&
\frac{(\ell-2\rho-1)\sinh(x)}{2}
\frac{P_{\ell-1}^{(1/2+\epsilon-\rho,-1/2-\epsilon-\rho)}(\cosh(x))}
{P_{\ell}^{(-1/2+\epsilon-\rho,-3/2-\epsilon-\rho)}(\cosh(x))}+f(x;J_k)\nonumber\\
W_{1-}(x;\epsilon,\rho,\ell)&=&
\frac{(\ell-2\rho-1)\sinh(x)}{2}
\frac{P_{\ell-1}^{(-1/2+\epsilon-\rho,1/2-\epsilon-\rho)}(\cosh(x))}
{P_{\ell}^{(-3/2+\epsilon-\rho,-1/2-\epsilon-\rho)}(\cosh(x))}+f(x;J_k)\nonumber
\end{eqnarray}
where $\epsilon=M+\beta_jI_j$, $\rho=d_jI_j$, $M=\frac{1}{n}\sum_{i=1}^n m_i$, $\beta_j,d_j$ are constants and $I_j,J_k$ are invariants in the $m_i$ under the change $m_i\rightarrow m_i-1$, $i=1,\dots,n$.

\item We have
\begin{eqnarray}
W_0(x;\epsilon,\rho,\ell)&=&2(\ell+\epsilon)\coth(2x)+2\rho\csch(2x)\nonumber\\
W_{1+}(x;\epsilon,\rho,\ell)&=&
-(2\rho-\ell+1)\sinh(2x)
\frac{P_{\ell-1}^{(-1/2-\ell-\epsilon-\rho,1/2+\ell+\epsilon-\rho)}(\cosh(2x))}
{P_{\ell}^{(-3/2-\ell-\epsilon-\rho,-1/2+\ell+\epsilon-\rho)}(\cosh(2x))}+f(x;J_k)\nonumber\\
W_{1-}(x;\epsilon,\rho,\ell)&=&
-(2\rho-\ell+1)\sinh(2x)
\frac{P_{\ell-1}^{(1/2-\ell-\epsilon-\rho,-1/2+\ell+\epsilon-\rho)}(\cosh(2x))}
{P_{\ell}^{(-1/2-\ell-\epsilon-\rho,-3/2+\ell+\epsilon-\rho)}(\cosh(2x))}+f(x;J_k)\nonumber
\end{eqnarray}
where $\epsilon=M+\beta_jI_j$, $\rho=d_jI_j$, $M=\frac{1}{n}\sum_{i=1}^n m_i$, $\beta_j,d_j$ are constants and $I_j,J_k$ are invariants in the $m_i$ under the change $m_i\rightarrow m_i-1$, $i=1,\dots,n$.

\item We have
\begin{eqnarray}
W_0(x;\epsilon,\rho)&=&\frac{\epsilon}{x}+\rho x\nonumber\\
W_{1+}(x;\epsilon,\rho)&=&-\frac{4\rho x}{2\epsilon+1-2\rho x^2}+f(x;J_k)\nonumber\\
W_{1-}(x;\epsilon,\rho)&=&-\frac{4\rho x}{2\epsilon-1-2\rho x^2}+f(x;J_k)\nonumber\\
\end{eqnarray}
where $\epsilon=M+d_jI_j$, $\rho=\frac{1}{2}\beta_jI_j$, $M=\frac{1}{n}\sum_{i=1}^n m_i$, $\beta_j,d_j$ are constants and $I_j,J_k$ are invariants in the $m_i$ under the change $m_i\rightarrow m_i-1$, $i=1,\dots,n$.

\item We have
\begin{eqnarray}
W_0(x;\epsilon,\rho,\ell)&=&\frac{\epsilon}{x}+\rho x\nonumber\\
W_{1+}(x;\epsilon,\rho,\ell)&=&2 x \rho
\frac{L_{\ell-1}^{(-1/2-\epsilon)}(-\rho x^2)}{L_{\ell}^{(-3/2-\epsilon)}(-\rho x^2)}+f(x;J_k)\nonumber\\
W_{1-}(x;\epsilon,\rho,\ell)&=&
2 x \rho
\frac{L_{\ell-1}^{(1/2-\epsilon)}(-\rho x^2)}{L_{\ell}^{(-1/2-\epsilon)}(-\rho x^2)}+f(x;J_k)\nonumber\\
\end{eqnarray}
where $\epsilon=M+d_jI_j$, $\rho=\frac{1}{2}\beta_jI_j$, $M=\frac{1}{n}\sum_{i=1}^n m_i$, $\beta_j,d_j$ are constants and $I_j,J_k$ are invariants in the $m_i$ under the change $m_i\rightarrow m_i-1$, $i=1,\dots,n$.

\item We have
\begin{eqnarray}
W_0(x;\epsilon,\ell)&=&\frac{\epsilon+\ell}{x}-x\nonumber\\
W_{1+}(x;\epsilon,\ell)&=&
2x\frac{L_{\ell-1}^{(1/2+\ell+\epsilon)}(-x^2)}
{L_{\ell}^{(-1/2+\ell+\epsilon)}(-x^2)}+f(x;J_k)\nonumber\\
W_{1-}(x;\epsilon,\ell)&=&
2x\frac{L_{\ell-1}^{(-1/2+\ell+\epsilon)}(-x^2)}
{L_{\ell}^{(-3/2+\ell+\epsilon)}(-x^2)}+f(x;J_k)\nonumber
\end{eqnarray}
where $\epsilon=M+d_jI_j$, $M=\frac{1}{n}\sum_{i=1}^n m_i$, $d_j$ are constants and $I_j,J_k$ are invariants in the $m_i$ under the change $m_i\rightarrow m_i-1$, $i=1,\dots,n$.

\item We have
\begin{eqnarray}
W_0(x;\epsilon,\ell)&=&
\frac{\epsilon+\ell}{x}-x\nonumber\\
W_{1+}(x;\epsilon,\ell)&=&
2\ell x
\frac{{}_1F_1\left(\begin{array}{c}1-\ell\\3/2+\ell+\epsilon\end{array}
  \Bigm|-x^2\right)}
  {(1/2+\ell+\epsilon){}_1F_1\left(\begin{array}{c}-\ell\\1/2+\ell+\epsilon\end{array}
  \Bigm|-x^2\right)}+f(x;J_k)\nonumber\\
W_{1-}(x;\epsilon,\ell)&=&
2\ell x
\frac{{}_1F_1\left(\begin{array}{c}1-\ell\\1/2+\ell+\epsilon\end{array}
  \Bigm|-x^2\right)}
  {(-1/2+\ell+\epsilon){}_1F_1\left(\begin{array}{c}-\ell\\-1/2+\ell+\epsilon\end{array}
  \Bigm|-x^2\right)}+f(x;J_k)\nonumber
\end{eqnarray}
where ${}_1F_1\left(\begin{array}{c}a\\b\end{array}
  \Bigm|x\right)$ denotes the confluent hypergeometric function,
$\epsilon=M+d_jI_j$, $M=\frac{1}{n}\sum_{i=1}^n m_i$, $d_j$ are constants and $I_j,J_k$ are invariants in the $m_i$ under the change $m_i\rightarrow m_i-1$, $i=1,\dots,n$.

\item We have
\begin{eqnarray}
W_0(x;\epsilon,\rho)&=&-\epsilon\tan(x)-\rho\sec(x)\nonumber\\
W_{1+}(x;\epsilon,\rho)&=&\frac{2\rho\cos(x)}{2\epsilon+1+2\rho\sin(x)}+f(x;J_k)\nonumber\\
W_{1-}(x;\epsilon,\rho)&=&\frac{2\rho\cos(x)}{2\epsilon-1+2\rho\sin(x)}+f(x;J_k)\nonumber\\
\end{eqnarray}
where $\epsilon=M-\beta_jI_j$, $\rho=d_jI_j$, $M=\frac{1}{n}\sum_{i=1}^n m_i$, $\beta_j,d_j$ are constants and $I_j,J_k$ are invariants in the $m_i$ under the change $m_i\rightarrow m_i-1$, $i=1,\dots,n$.

\item We have
\begin{eqnarray}
W_0(x;\epsilon,\rho,\ell)&=&2(\epsilon+\ell)\cot(2x)-2\rho\csc(2x)\nonumber\\
W_{1+}(x;\epsilon,\rho,\ell)&=&
-(2\rho+\ell-1)\sin(2x)
\frac{P_{\ell-1}^{(-1/2-\ell-\epsilon+\rho,1/2+\ell+\epsilon+\rho)}(\cos(2x))}
{P_{\ell}^{(-3/2-\ell-\epsilon+\rho,-1/2+\ell+\epsilon+\rho)}(\cos(2x))}+f(x;J_k)\nonumber\\
W_{1-}(x;\epsilon,\rho,\ell)&=&
-(2\rho+\ell-1)\sin(2x)
\frac{P_{\ell-1}^{(1/2-\ell-\epsilon+\rho,-1/2+\ell+\epsilon+\rho)}(\cos(2x))}
{P_{\ell}^{(-1/2-\ell-\epsilon+\rho,-3/2+\ell+\epsilon+\rho)}(\cos(2x))}+f(x;J_k)\nonumber
\end{eqnarray}
where $\epsilon=M-\beta_jI_j$, $\rho=d_jI_j$, $M=\frac{1}{n}\sum_{i=1}^n m_i$, $\beta_j,d_j$ are constants and $I_j,J_k$ are invariants in the $m_i$ under the change $m_i\rightarrow m_i-1$, $i=1,\dots,n$.

\item We have
\begin{eqnarray}
W_0(x;\epsilon,\rho,\ell)&=&2\epsilon\cot(2x)+2\rho\csc(2x)\nonumber\\
W_{1+}(x;\epsilon,\rho,\ell)&=&
-\ell(2\rho+\ell-1)\sin(2x)\frac{\Gamma(1/2+\epsilon+\rho)\,{}_2F_1\left(\begin{array}{c}1-\ell,\ell+2\rho\\3/2+\epsilon+\rho\end{array}\Bigm|\sin^2(x)\right)}
{\Gamma(3/2+\epsilon+\rho)\,{}_2F_1\left(\begin{array}{c}-\ell,-1+\ell+2\rho\\1/2+\epsilon+\rho\end{array}\Bigm|\sin^2(x)\right)}+f(x;J_k)\nonumber\\
W_{1-}(x;\epsilon,\rho,\ell)&=&
-\ell(2\rho+\ell-1)\sin(2x)\frac{\Gamma(-1/2+\epsilon+\rho)\,{}_2F_1\left(\begin{array}{c}1-\ell,\ell+2\rho\\1/2+\epsilon+\rho\end{array}\Bigm|\sin^2(x)\right)}
{\Gamma(1/2+\epsilon+\rho)\,{}_2F_1\left(\begin{array}{c}-\ell,-1+\ell+2\rho\\-1/2+\epsilon+\rho\end{array}\Bigm|\sin^2(x)\right)}+f(x;J_k)\nonumber
\end{eqnarray}
where ${}_2F_1\left(\begin{array}{c}a,b\\c\end{array}\Bigm|x\right)$
is the hypergeometric function,
$\epsilon=M-\beta_jI_j$, $\rho=d_jI_j$, $M=\frac{1}{n}\sum_{i=1}^n m_i$, $\beta_j,d_j$ are constants and $I_j,J_k$ are invariants in the $m_i$ under the change $m_i\rightarrow m_i-1$, $i=1,\dots,n$.

\item We have
\begin{eqnarray}
W_0(x;\epsilon,\rho,\ell)&=&\epsilon\tanh(x)
+i\rho\sech(x)\nonumber\\
W_{1+}(x;\epsilon,\rho,\ell)&=&
\frac{1}{2}i(\ell-2\rho-1)\cosh(x)
\frac{P_{\ell-1}^{(-\rho+\epsilon+1/2,-\rho-\epsilon-1/2)}(i\sinh(x))}
{P_\ell^{(-\rho+\epsilon-1/2,-\rho-\epsilon-3/2)}(i\sinh(x))}+f(x;J_k)\nonumber\\
W_{1-}(x;\epsilon,\rho,\ell)&=&
\frac{1}{2}i(\ell-2\rho-1)\cosh(x)
\frac{P_{\ell-1}^{(-\rho+\epsilon-1/2,-\rho-\epsilon+1/2)}(i\sinh(x))}
{P_\ell^{(-\rho+\epsilon-3/2,-\rho-\epsilon-1/2)}(i\sinh(x))}+f(x;J_k)\nonumber
\end{eqnarray}
where $\epsilon=M+\beta_jI_j$, $\rho=d_jI_j$, $M=\frac{1}{n}\sum_{i=1}^n m_i$, $\beta_j,d_j$ are constants and $I_j,J_k$ are invariants in the $m_i$ under the change $m_i\rightarrow m_i-1$, $i=1,\dots,n$.

\end{enumerate}

The group theory approach of \cite{RamBagKhaKumManYad17} can be achieved easily substituting there by our current quantities $F(x)\rightarrow G(x)$, $G(x)\rightarrow I_j v_j$, $a\rightarrow\alpha$, $b\rightarrow \beta_jI_j$,
$U(x;m)\rightarrow W_{1+}(x;\epsilon,\rho)-W_{1-}(x;\epsilon,\rho)$, $U(x;m)\rightarrow W_{1+}(x;\epsilon,\rho,\ell)-W_{1-}(x;\epsilon,\rho,\ell)$ or $U(x;m)\rightarrow W_{1+}(x;\epsilon,\ell)-W_{1-}(x;\epsilon,\ell)$,
depending on the case.

\section{More generalizations}\label{moregen}

In \cite{InfHul51,CarRam00c} there are solutions to the shape invariance condition (\ref{sicond}) for only one parameter $m$ subject to $m\rightarrow m-1$ of the form
$$
k(x;m)=\frac{q}{m}+m\,k_1(x)
$$
where $q$ is a constant. The function $k_1(x)$ must satisfy again the differential equation (\ref{eqRic}) or (\ref{eqalpha}),
whose solutions have been discussed earlier in this paper.

We propose the following generalization to $m_1,m_2,\dots,m_n$ parameters subject to translation $m_i\rightarrow m_i-1$, $i=1,\dots,n$:
$$
k(x;\epsilon,\rho)=\frac{\rho}{\epsilon}+\epsilon\, G(x)
$$
where $\epsilon=M+d_jI_j=\frac{1}{n}\sum_{i=1}^n m_i+d_jI_j$, $d_j$ are constants, $I_j$ are invariants as before, $\rho=I$ is another invariant and $G(x)$ is a solution of (\ref{eqRic}).
The non-constant solutions for $G(x)$ can be reduced to the basic forms (\ref{soltanh}), (\ref{solcoth}), (\ref{Gcoul}), (\ref{soltan}) and (\ref{solcot})

Let us describe in what follows the corresponding superpotential, partner potentials, remainder of the shape invariance condition, and eigenenergies and normalized eigenstates of the potential $V(x;\epsilon,\rho)$ for each case.

For these cases no rational extensions are known, nor the formulation in group theoretical terms.

\subsection{Case (\ref{soltanh}) (Rosen-Morse 2 type)}

We have set $\epsilon=M+d_jI_j$, $\rho=I$.
We have, for $\epsilon>\rho/\epsilon$ and $\epsilon+\rho/\epsilon>0$,
\begin{eqnarray}
k(x;\epsilon,\rho)&=&\epsilon\tanh(x)+\frac{\rho}{\epsilon}\,\quad x\in(-\infty,\infty)\nonumber\\
V(x;\epsilon,\rho)&=&\epsilon^2\tanh^2(x)+2\rho\tanh(x)-\epsilon\sech^2(x)+\rho^2/\epsilon^2\nonumber\\
\tilde{V}(x;\epsilon,\rho)&=&\epsilon^2\tanh^2(x)+2\rho\tanh(x)+\epsilon\sech^2(x)+\rho^2/\epsilon^2\nonumber\\
R(\epsilon,\rho)&=&1+2\epsilon-\frac{(2\epsilon+1)\rho^2}{\epsilon^2(\epsilon+1)^2}\nonumber\\
E_k&=&k(k-2\epsilon)\left(\frac{\rho^2}{(k-\epsilon)^2\epsilon^2}-1\right)\nonumber\\
\zeta_k(x;\epsilon,\rho)&=&
2^{1/2+k-\epsilon}k!\sqrt{\frac{\Gamma(2\epsilon-2k)}
{\Gamma\left(\epsilon-k+\frac{\rho}{k-\epsilon}\right)
\Gamma\left(\epsilon-k-\frac{\rho}{k-\epsilon}\right)}}
e_k(\epsilon,\rho)\nonumber\\
&&(1-\tanh(x))^{\frac{1}{2}\left(\epsilon-k+\rho/(\epsilon-k)\right)}
(1+\tanh(x))^{\frac{1}{2}\left(\epsilon-k-\rho/(\epsilon-k)\right)}\nonumber\\
&&P_k^{(\epsilon-k+\rho/(\epsilon-k),\epsilon-k-\rho/(\epsilon-k))}(\tanh(x))\nonumber
\end{eqnarray}
where
\begin{equation}
e_k(\epsilon,\rho)=\left\{\begin{array}{ll}
1, & k=0\\
\frac{(2\epsilon-k)e_{k-1}(\epsilon-1,\rho)}
{\epsilon\sqrt{k(2\epsilon-k)-\rho^2/(k-\epsilon)^2+\rho^2/\epsilon^2}}, & k>0
\end{array}\right.\label{defe}
\end{equation}

\subsection{Case (\ref{solcoth}) (Eckart type)}

We have set $\epsilon=M+d_jI_j$, $\rho=I$.
We have, for $\epsilon<\frac{1}{2}$ and  $\epsilon+\rho/\epsilon>0$,
\begin{eqnarray}
k(x;\epsilon,\rho)&=&\epsilon\coth(x)+\frac{\rho}{\epsilon},
\quad x\in(0,\infty)\nonumber\\
V(x;\epsilon,\rho)&=&\epsilon^2\coth^2(x)+2\rho\coth(x)+\epsilon\csch^2(x)+\rho^2/\epsilon^2\nonumber\\
\tilde{V}(x;\epsilon,\rho)&=&\epsilon^2\coth^2(x)+2\rho\coth(x)-\epsilon\csch^2(x)+\rho^2/\epsilon^2\nonumber\\
R(\epsilon,\rho)&=&1+2\epsilon-\frac{(2\epsilon+1)\rho^2}{\epsilon^2(\epsilon+1)^2}\nonumber\\
E_k&=&k(k-2\epsilon)\left(\frac{\rho^2}{(k-\epsilon)^2\epsilon^2}-1\right)\nonumber\\
\zeta_k(x;\epsilon,\rho)&=&
2^{1/2+k-\epsilon}k!\sqrt{\frac{\Gamma(1+k-\epsilon+\rho/(\epsilon-k))}
{\Gamma\left(1+2k-2\epsilon\right)
\Gamma\left(\epsilon-k-\frac{\rho}{k-\epsilon}\right)}}
e_k(\epsilon,\rho)\nonumber\\
&&(\coth(x)-1)^{\frac{1}{2}\left(\epsilon-k+\rho/(\epsilon-k)\right)}
(\coth(x)+1)^{\frac{1}{2}\left(\epsilon-k-\rho/(\epsilon-k)\right)}\nonumber\\
&&P_k^{(\epsilon-k+\rho/(\epsilon-k),\epsilon-k-\rho/(\epsilon-k))}(\coth(x))\nonumber
\end{eqnarray}
where $e_k(\epsilon,\rho)$ is given by (\ref{defe}).

\subsection{Case (\ref{Gcoul}) (Coulomb type)}

We have set $\epsilon=M+d_jI_j$, $\rho=I$.
We have, for $\epsilon\neq 0$, $\epsilon<1/2$, $\rho/\epsilon>0$,
\begin{eqnarray}
k(x;\epsilon,\rho)&=&\frac{\epsilon}{x}+\frac{\rho}{\epsilon},
\quad x\in(0,\infty)\nonumber\\
V(x;\epsilon,\rho)&=&\frac{2\rho}{x}+\frac{\epsilon(\epsilon+1)}{x^2}
+\frac{\rho^2}{\epsilon^2}\nonumber\\
\tilde{V}(x;\epsilon,\rho)&=&\frac{2\rho}{x}+\frac{\epsilon(\epsilon-1)}{x^2}
+\frac{\rho^2}{\epsilon^2}\nonumber\\
R(\epsilon,\rho)&=&-\frac{(2\epsilon+1)\rho^2}{\epsilon^2(\epsilon+1)^2}\nonumber\\
E_k&=&k(k-2\epsilon)\frac{\rho^2}{(k-\epsilon)^2\epsilon^2}\nonumber\\
\zeta_k(x;\epsilon,\rho)&=&
(-1)^k k!
\sqrt{-\frac{(k-\epsilon)^2}{\rho}
\left(\frac{\rho}{\epsilon-k}\right)^{2\epsilon-2k}
\Gamma(2k-2\epsilon)}p_k(\epsilon,\rho)
(2x)^{-\epsilon}\exp\left(\rho x/(k-\epsilon)\right)L_k^{(-1-2\epsilon)\left(\frac{2\rho x}{\epsilon-k}\right)}\nonumber
\end{eqnarray}
where
\begin{equation}
p_k(\epsilon,\rho)=\left\{\begin{array}{ll}
1, & k=0\\
\frac{(2\epsilon-k)}{\epsilon}\sqrt{\frac{(k-\epsilon)^2\epsilon^2}{k(k-2\epsilon)\rho^2}}
p_{k-1}(\epsilon-1,\rho), & k>0
\end{array}\right.
\end{equation}

\subsection{Case (\ref{soltan}) (Rosen-Morse 1 type)}

We have set $\epsilon=M+d_jI_j$, $\rho=I$.
We have, for $\epsilon<1/2$,
\begin{eqnarray}
k(x;\epsilon,\rho)&=&-\epsilon\tan(x)+\rho/\epsilon,
\quad x\in(-\pi/2,\pi/2)\nonumber\\
V(x;\epsilon,\rho)&=&\epsilon^2\tan^2(x)-2\rho\tan(x)+\epsilon\sec^2(x)+\rho^2/\epsilon^2\nonumber\\
\tilde{V}(x;\epsilon,\rho)&=&\epsilon^2\tan^2(x)-2\rho\tan(x)-\epsilon\sec^2(x)+\rho^2/\epsilon^2\nonumber\\
R(\epsilon,\rho)&=&-1-2\epsilon-\frac{(2\epsilon+1)\rho^2}{\epsilon^2(\epsilon+1)^2}\nonumber\\
E_k&=&k(k-2\epsilon)\left(\frac{\rho^2}{(k-\epsilon)^2\epsilon^2}+1\right)\nonumber\\
\zeta_k(x;\epsilon,\rho)&=&(-i)^kk!
\frac{\left|\Gamma
\left(1+k-\epsilon+\frac{i\rho}{k-\epsilon}\right)\right|}
{\sqrt{\pi \Gamma(1+2k-2\epsilon)}}
u_k(\epsilon,\rho)(2\cos(x))^{k-\epsilon}
\exp\left(\frac{\rho x}{k-\epsilon}\right)\nonumber\\
&&P_k^{(\epsilon-k+i\rho/(\epsilon-k),
\epsilon-k-i\rho/(\epsilon-k))}(-i\tan(x))\nonumber
\end{eqnarray}
where
\begin{equation}
u_k(\epsilon,\rho)=\left\{\begin{array}{ll}
1, & k=0\\
\frac{(2\epsilon-k)u_{k-1}(\epsilon-1,\rho)}
{\epsilon\sqrt{k(k-2\epsilon)-\rho^2/(k-\epsilon)^2
+\rho^2/\epsilon^2}}, & k>0
\end{array}\right.
\label{defp}
\end{equation}

\subsection{Case (\ref{solcot})}

We have set $\epsilon=M+d_jI_j$, $\rho=I$.
Finally, we have, for $\epsilon<1/2$,
\begin{eqnarray}
k(x;\epsilon,\rho)&=&\epsilon\cot(x)+\rho/\epsilon,\quad x\in(0,\pi)\nonumber\\
V(x;\epsilon,\rho)&=&\epsilon^2\cot^2(x)+2\rho\cot(x)+\epsilon\csc^2(x)
+\rho^2/\epsilon^2\nonumber\\
\tilde{V}(x;\epsilon,\rho)&=&\epsilon^2\cot^2(x)+2\rho\cot(x)-\epsilon\csc^2(x)
+\rho^2/\epsilon^2\nonumber\\
R(\epsilon,\rho)&=&-1-2\epsilon-\frac{(2\epsilon+1)\rho^2}{\epsilon^2(\epsilon+1)^2}\nonumber\\
E_k&=&k(k-2\epsilon)\left(\frac{\rho^2}{(k-\epsilon)^2\epsilon^2}+1\right)\nonumber\\
\zeta_k(x;\epsilon,\rho)&=&(-i)^k k!
\frac{\left|\Gamma
\left(1+k-\epsilon+\frac{i\rho}{k-\epsilon}\right)\right|}
{\sqrt{\pi \Gamma(1+2k-2\epsilon)}}
u_k(\epsilon,\rho)(2\sin(x))^{k-\epsilon}
\exp\left(\frac{\rho (2x-\pi)}{2(k-\epsilon)}\right)\nonumber\\
&&P_k^{(\epsilon-k+i\rho/(\epsilon-k),
\epsilon-k-i\rho/(\epsilon-k))}(i\cot(x))\nonumber
\end{eqnarray}
where $u_k(\epsilon,\rho)$ is given again by (\ref{defp}).

\section{Conclusions}\label{disccon}

We have found in this paper a way of generalizing the previously known cases of shape invariant potentials (and their rational extensions) to the inclusion of an arbitrary number $r$ of quantities that are invariant under the change of $n\geq1$ parameters subject to translation. When $n=1$ the invariants are essentially periodic functions in only one parameter, with period $1$, and have not been observed before to the best of our knowledge. When $n\geq1$ the non-trivial invariants may enter in the expression of quantities of physical significance, as it is for example the spectrum of the problem. This means that the spectrum and perhaps other meaningful physical quantities could be engineered to a great extent. The natural consequence is that this opens the door to a possible multitude of quantum applications, like for example quantum computing and physics of bilayer graphene \cite{FerGarCam20}.
Also, other fundamental questions might be relevant. For example, how a modification of the approach in \cite{BouGanMal11} could generate the new solutions found here. Also, whether these new found cases satisfy the SWKB formalism \cite{GanMalRasBou20} or not.
All these would be probably interesting questions for future research.

\section*{Author contributions}
Arturo Ramos: Conceptualization, formal analysis, funding acquisition, investigation, methodology, resources, software, validation, visualization, writing-original draft, writing-review \& editing.

\section*{Competing interests statement}
The author declares to have no competing interests concerning the research carried out in this article.





\section*{Acknowledgments}
The work of Arturo Ramos has been supported by the Spanish \emph{Ministerio de Econom\'{\i}a y Competitividad} (ECO2017-82246-P) and by Aragon Government (ADETRE Reference Group).


\end{document}